\documentclass{article}

\def\beq{\begin{equation}}
\def\eeq{\end{equation}}
\def\pd{\partial}
\def\bea{\begin{eqnarray}}
\def\eea{\end{eqnarray}}

\renewcommand{\thefootnote}{\fnsymbol{footnote}}
\begin{document}
\vspace{0.6cm}
\begin{center}
{\Large{\bf A note on the supersymplectic structure of triplectic formalism}}\\
\vspace{1cm}

{\large Bodo Geyer}$^{a}$\footnote{e-mail: geyer@itp.uni-leipzig.de},
{\large Peter Lavrov}$^{a, b}$\footnote{e-mail: lavrov@tspu.edu.ru} and
{\large Armen Nersessian $^{a,c,d}$}
\footnote{e-mail: nerses@thsun1.jinr.ru}

\vspace{0.5cm}
{\normalsize\it $^{a)}$ Center of Theoretical Studies, Leipzig University,
Augustusplatz 10/11, D-04109 Leipzig, Germany}\\
{\normalsize\it $^{b)}$ Tomsk State Pedagogical University,
634041 Tomsk, Russia}\\
{\normalsize\it $^{c)}$ Yerevan State University, A. Manoogian St.,1 ,
Yerevan, 375025, Armenia}\\
{\normalsize\it $^{d)}$ Yerevan Physics Institute, Alikhanian Brothers St.,2 , Yerevan, 375036, Armenia}
\end{center}
\smallskip
\begin{abstract}
We equip the whole space of fields of the triplectic formalism of Lagrangian
quantization with an even supersymplectic structure and clarify its geometric meaning. 
We also discuss its relation to a closed two-form arising naturally in the
superfield approach to the triplectic formalism.

\end{abstract}
\bigskip
\setcounter{page}1
\renewcommand{\thefootnote}{\arabic{footnote}}
\setcounter{footnote}0
\setcounter{equation}0

\section{Introduction}

The Batalin-Vilkovisky (BV) formalism \cite{BV} of Lagrangian
quantization of general gauge theories, since its introduction,
attracts permanent interest due to its covariance, universality
and mathematical elegance. Now, its area of physical applications is
much wider than offered in the initial prescription. The BV-formalism
is outstanding also from the mathematical point of view because it
is formulated in terms of seemingly exotic objects: the antibracket
(odd Poisson bracket) and the related second-order operator $\Delta$.

The study of the geometric structure of the BV formalism allowed
to introduce its interpretation in terms of more traditional
mathematical objects \cite{bvgeom}. On the other hand, there
exists a more complicated, $Sp(2)$ symmetric extension of the BV
formalism \cite{BLT}, and of its geometrized version known as
`triplectic formalism' \cite{BM,BMS} (see also \cite{GGL}).

In the BV formalism the original set of `physical' fields $x^i,
\epsilon(x^i)\equiv \epsilon_i$ (including ghosts, antighosts,
Lagrangian multipliers etc.), is doubled by the `antifields'
$\theta_{i}$ with opposite grading. On this set of fields and
antifields the {\it nondegenerate} antibracket and the
corresponding $\Delta-$operator are defined. Differently, the
triplectic formalism deals with two sets of auxiliary fields
$\theta_{ai},\, a = 1,2$, which could be arranged in the set of
triplets $(x^i,\theta_{ai}), \epsilon({\theta}_{ai})=\epsilon
_i+1$. The fields $x^i$ parametrize the subspace ${\cal M}_0$
endowed with an {\em even} (super)symplectic structure, \beq
\label{w}
 \omega = \omega_{ij}(x)\;dx^j\wedge dx^i,
 \qquad
 d\omega =\omega_{ij,k}\;dx^k\wedge dx^j\wedge dx^i=0\,.
\eeq
The components $\omega_{ij}$ obey the relations
\beq
\omega_{ij}=-(-1)^{\epsilon_i\epsilon_j}\;\omega_{ji}\,,
 \qquad (-1)^{\epsilon_j\epsilon_k}\omega_{ki,j}+
 (-1)^{\epsilon_i\epsilon_j}\omega_{jk,i}+
 (-1)^{\epsilon_i\epsilon_j}\omega_{jk,i}= 0\,,
 \qquad
 \epsilon(\omega_{ij})=\epsilon_i+\epsilon_j\,.
\eeq
The inverse tensor $\omega^{ij},\,$
$\omega^{ik}\;\omega_{kj}(-1)^{\epsilon_k}=\delta^i_j$,\,
defines on ${\cal M}_0$ an even Poisson bracket,
\beq
 \label{pbo}
 \left\{ f(x),g(x)\right\}_0
 =\frac{\partial_r f}{\partial x^i}\;
 \omega^{ij}\; \frac{\partial_l g}{\partial x^j}\,.
\eeq

The whole space of fields and antifields, ${\cal M}$,
is equipped with a pair of {\em degenerate} antibrackets,
\beq\label{ab}
 \left( f(x, \theta) ,g(x, \theta )\right)^a
 =
 \frac{\partial_l f}{\partial x^i}\;
 \frac{\partial_r g}{\partial \theta_{ai}}\,
 -\, \frac{\partial_l f}{\partial \theta_{ai}}\;
 \frac{\partial_r g}{\partial x^{i}}\,,
\eeq
together with a related pair of operators $\Delta^a$;
also some additional odd vector fields $V^a$ are needed which,
in some special case \cite{BM}, could be absorbed by the action.

The geometry underlying the triplectic formalism, is quite rich
and unusual. There were some efforts to understand it from various
points of view \cite{grig}, as well as to find explicit non-trivial
examples of such triplectic spaces ${\cal M}$. In particular,
in our previous papers \cite{gln,GL03}, we tried to give a covariant
(coordinate-free) realization of the triplectic formalism, equipping
the space ${\cal M}_0$ with a connection which respects the symplectic
structure (\ref{w}). However, we found that the basic relations of the
triplectic formalism could be fulfilled in such an approach for flat
symplectic connections only. On the other hand, in \cite{gln} we found 
that the
implementation of that symplectic connection allows to equip the whole
triplectic space ${\cal M}$ with an even symplectic structure which
provides the triplectic formalism with a well-defined integration measure.

In the papers \cite{gln}, considering an even symplectic structure, we
restricted ourselves to the case of flat connections. We also assumed,
for the sake of simplicity, that the initial space ${\cal M}_0$ is a
purely bosonic one. Of course, in the triplectic formalism, ${\cal M}_0$
is necessarily a supermanifold (containing, besides the original gauge
fields, also ghosts and antighosts as well as matter fields). In the
paper \cite{GL03} we repaired this, but gave no detailed study of the 
(even) supersymplectic structure and its geometric implications.

In this note, avoiding the just mentioned restrictions, we
equip the whole space ${\cal M}$ of the triplectic formalism with
an even supersymplectic structure $\Omega$ and we clarify their
geometric origin. With the aim to lay the ground for finding a 
realization of the triplectic algebra also in that general non-flat
case we consider a closed two-form arising naturally in the superfield
formulation of triplectic formalism \cite{l} and try to relate it to
$\Omega$. This, however, seems not to be straightforward and should be
re-considered.

\section{Even (Super) Symplectic Structures}

First, we repeat the introduction of an even symplectic structure
in the restricted case of a (bosonic) manifold ${\cal M}_0$ but without
assuming its flatness. After that, we give a concise geometric
formulation of constructing that symplectic structure which, then,
will be generalized to the case of a supermanifold ${\cal M}_0$.

When the Poisson bracket (\ref{pbo}) is non-degenerate, the
superspace ${\cal M}$ can be equipped with both an {\it even
symplectic structure} and a corresponding {\it non-degenerate
Poisson bracket} analogous to \cite{gln},
 \beq \Omega
 =\omega+\alpha^{-1}\,d\!\left(\theta^{ia}\omega_{ij} D\theta_a^j\right)
 =\left(\omega_{ij}+\hbox{\large$\frac{1}{2\alpha}$} \, 
  R_{ijkl}\,\theta^{ka}\theta^{l}_a\right)\, dx^i\wedge dx^j
 + \hbox{\large$\frac{1}{\alpha}$} \,
  \omega_{ij}\,D\theta^{ia}\wedge D\theta_a^j\,,
 \label{ss0}
\eeq
where $\alpha$ is an arbitrary constant, the $Sp(2)-$indices are
lowered by the invariant $Sp(2)-$tensor $\epsilon_{ab}$,
$\theta^i_a = \epsilon_{ab}\,\theta^{ib}$, and the covariant derivative is
defined by
$D\theta^{ia}=d\theta^{ia}+\Gamma^i_{\,kl}\,\theta^a_{k}\,dx^l$;
thereby, $\Gamma^i_{\,kl}(x)$ are the coefficients of the connection which
respects the symplectic structure: 
\beq
\pd_k\omega_{ij}-\Gamma^l_{\,ki}\,\omega_{lj}-\omega_{il}\Gamma^l_{\,kj} =0\,,
\eeq
while $R_{ijkl}=\omega_{im}R^m_{~jkl}$, with $R^i_{jkl}$ being the curvature
components of that symplectic connection,
\begin{eqnarray}
\label{Rsp1}
R^i_{\,jkl}=
 -\,\Gamma^i_{\,kj,l}
 +\,\Gamma^i_{\,lj,k}
 +\,\Gamma^i_{\,km}\Gamma^m_{~lj}
 -\,\Gamma^i_{\,lm}\Gamma^m_{~kj}\,.
\end{eqnarray}
The indices $i$ are lowered by the help of the symplectic structure, e.g.
$\theta_{ia}=\omega_{ij}\theta^j_a\,.$

Obviously, the suggested symplectic structure $\Omega$ transforms covariant
under the following change of coordinates,
 \beq
 \label{trans}
 {\bar x}^i
 ={\bar x}^i(x),\quad {\bar\theta}^i_a
 =\frac{\partial{\bar x}^i}{\partial x^j}\theta^j_a,
 \eeq
so that $\theta^i_a$ could be identified with (two different) one-forms
$dx^i$. It is clear that, due to the presence of the $Sp(2)$
indices, it is possible to describe not only external forms on
${\cal M}_0$, i.e. antisymmetric covariant tensors, but also
specific symmetric ones as well. As is well-known, one could equip the
space ${\cal M}$ with a pair of antibrackets (\ref{ab}) which transform covariant under the coordinate changes (\ref{trans}).

Furthermore, using the symplectic structure (\ref{ss0}) we can introduce
the following nondegenerate Poisson bracket, thereby extending the Poisson bracket
(\ref{ab}) from ${\cal M}_0$ to the whole space ${\cal M}$,
\beq
 \{f(z), g(z)\}
 =(\nabla_i f)\,{\widetilde\omega}^{ij}( \nabla_j g)
 + \alpha\,\frac{\partial_r f}{\partial \theta_a^{i}}\,
 \omega^{ij}\frac{\partial_l g}{\partial \theta^{ja}}\,;
\label{ss}
\eeq
here, we used the notations
\begin{eqnarray}
 {\widetilde\omega}^{im}
 {\widetilde\omega}_{mj}
 &\equiv&
 {\widetilde\omega}^{im}\big(\omega_{mj}
 +
 \hbox{\large$\frac{1}{2\alpha}$}\,
 R_{mjkl}\,\theta^{ka}\,\theta^{l}_a\big)
 =\delta^i_j\,,
 \\
 \nabla_i
 &\equiv&
 \frac{\partial}{\partial x^i}
 - \Gamma^k_{\,ij}(x)\,\theta^{ja}\frac{\partial}{\partial\theta^{ka}}
 \quad {\rm with}\quad
 [\nabla_i,\nabla_j]
 =R^k_{\;lij}\;\theta^{la}\frac{\partial}{\partial\theta^{k}_{a}}\,.
\end{eqnarray}
Obviously, the operator $\nabla_k$ acts on the monomials
$\alpha_{a_1 \ldots a_n}=
\alpha_{[i_1 \ldots i_n]}\,\theta^{i_1}_{a_1}\ldots\theta^{i_n}_{a_n}$
as a covariant derivative
\beq
 \nabla_k \alpha_{a_1 \ldots a_n}
 =
 \alpha_{[i_1...i_n\,;\,k]}\,\theta^{i_1}_{a_1}\ldots\theta^{i_n}_{a_n}.
\eeq

The symplectic structure $\Omega$ which we introduced above has a
simple geometrical meaning. To show this, let us remind some
general procedure for introducing a supersymplectic structure:
\medskip

(1) Let us consider some supermanifold ${\cal M}$ being given
as the vector bundle of some symplectic manifold ${\cal M}_0$.
Furthermore, let $\theta^\mu$ be odd coordinates parametrizing the
fibers of that bundle, and let $x^i$ be local coordinates of the
base manifold ${\cal M }_0$. Let $g_{\mu\nu}=g_{\mu\nu}(x)$ be a
metric on the bundle, and $\Gamma^\mu_{\;i\nu}$ be the components
of its connection, so that
\beq
 g_{\mu\nu} = g_{\nu\mu},
 \qquad
 g_{\mu\nu;k} = g_{\mu\nu,k} - g_{\mu\alpha}\Gamma^{\alpha}_{k\nu} -
 g_{\alpha\nu}\Gamma^{\alpha}_{k\mu} = 0\,.
 \label{gG}
\eeq
On such a supermanifold we can define a symplectic structure as follows:
\begin{eqnarray}
 \label{tO}
 \Omega
 &=&\omega+
 \hbox{\large$\frac{1}{\alpha}$}
 d\left(\theta^{\mu}g_{\mu\nu}(x){\cal D}\theta^{\nu}\right)
 \nonumber\\
 &=&\omega + \hbox{\large$\frac{1}{2\alpha}$}
 R_{\nu \mu ki}\theta^{\nu}\theta^{\mu}dx^i\wedge dx^k
 +
 \hbox{\large$\frac{1}{\alpha}$}g_{\mu\nu}{\cal D}
 \theta^{\nu} \wedge {\cal D}\theta^{\mu}\,,
\end{eqnarray}
where
${\cal D}\theta^\mu=d\theta^\mu+\Gamma^\mu_{\;\nu i}\,\theta^\nu dx^i$,
$R_{\mu \nu ki}=g_{\mu\alpha}R^{\alpha}_{\;\nu ki}$, while
$R^{\mu}_{\;\nu ki}$ are the curvature components of the connection.

Now, let us specify this symplectic structure to our case, i.e.,
let us choose $\theta^\mu=\theta^{ia}$. In this specification
$\mu,\nu$ are multi-indices: $\mu=(i,a)$, $\nu=(j,b)$, and we
choose the following metric and connection:
 \beq
 g_{\mu\nu}=\omega_{ij}\epsilon_{ab}\,,
 \qquad
 \Gamma^\mu_{\;\nu i} = \Gamma^i_{\,jk}\,\delta^a_b\,.
 \label{gGamma}
 \eeq
Upon such specification, from the covariant constancy of the
metric $g_{\mu\nu}$, Eq.~(\ref{gG}), it immediately follows that
$\Gamma^i_{\,jk}$ is a symplectic connection on ${\cal M}_0$. The
curvature of that connection is also reduced to the curvature of
the symplectic connection,
 \beq
 \label{Rsp}
 R_{\mu\nu kl}
 =g_{\mu\alpha}\,R^{\alpha}_{\,\nu kl}
 =\epsilon_{ab}\,\omega_{im}\,R^m_{\;jkl}
 =\epsilon_{ab}\,R_{ijkl}\,,
 \qquad
 R^i_{j\,kl}
 =-\Gamma^i_{\,kj,l}
  +\Gamma^i_{\,lj,k}
  +\Gamma^i_{\,km}\Gamma^m_{\;lj}
  -\Gamma^i_{\,lm}\Gamma^m_{\;kj}\,.
\eeq
Hence, we get precise correspondence with the symplectic structure
(\ref{ss0}).
\medskip

(2) It is easy to extend the above construction to the case when
${\cal M}_0$ is an even symplectic {\em supermanifold} with local
coordinates $x^i$, $\epsilon(x^i)\equiv \epsilon_i$. We shall
follow De Witt's definitions and conventions concerning tensor
fields on supermanifolds \cite{DeWitt} (see also \cite{GL03}. 
In particular, if the sets
$\{e_i=\frac{\partial_r }{\partial x^i}\}$ and $\{e^i=dx^i\}$ are
coordinate bases in the tangent and the cotangent spaces,
respectively, then they transform under a change of local
coordinates $x^i\rightarrow {\bar x}^i={\bar x}^i(x)$ according to
the rules
\begin{eqnarray}
\label{lb}
 {\bar e}_i=e_j\frac{\partial_r x^j}{\partial {\bar x}^i}\,,
 \qquad
 {\bar e}^i=e^j\frac{\partial{\bar x^i}}{\partial x^j}\,.
\end{eqnarray}

These vectors are dual with respect to an inner product operation,
$\langle\,\cdot\,,\,\cdot\,\rangle$,
\begin{eqnarray}
 \label{lr}
 \langle e^i,e_j \rangle \;=\;\delta^i_j\,,
 \qquad
 \langle e_j,e^i\rangle\;=\;(-1)^{\epsilon_i}\delta^i_j\,,
\end{eqnarray}
obeying the following properties:
\begin{eqnarray}
\label{lr1}
 \langle \omega,X_1+X_2\rangle\;
 =\;\langle\omega,X_1\rangle\;+\,\langle\omega,X_2\rangle,
 \qquad
 \langle\omega,X\rangle\;=\;\langle X,\omega\rangle\;(-1)^{\epsilon(\omega)\epsilon(X)},
\end{eqnarray}
and
\begin{eqnarray}
\label{lr2}
 \langle\omega,X_1\,X_2\rangle\;
 =\;\langle\omega,X_1\rangle\,X_2\,+\,\langle\omega,X_2\rangle\,X_1\,
 (-1)^{\epsilon(X_1)\epsilon(X_2)}\,.
\end{eqnarray}

The coordinates parametrising the fibers,
$\theta^\mu=\theta^{ia}$, could also be even and odd:
$\epsilon(\theta^\mu)=\epsilon_\mu +1=\epsilon_i+1$. In that case
the analog of the supersymplectic structure (\ref{tO}) reads
\begin{eqnarray}
\label{A}
 {\Omega}
 &=&\omega+ \hbox{\large$\frac{1}{\alpha}$} d\left(g_{\mu\nu}(x)\theta^{\nu}
 {\cal D}\theta^{\mu}(-1)^{\epsilon_{\nu}}\right)
 \nonumber\\
 &=&\omega
 - \hbox{\large $\frac{1}{2\alpha}$}
 g_{\mu\alpha}R^\alpha_{~\nu ki}
 dx^i\wedge dx^k\theta^{\nu}\theta^{\mu}(-1)^{\epsilon_{\nu}}+
 \hbox{\large$\frac{1}{\alpha}$}
 g_{\mu\nu}{\cal D}\theta^{\nu}\wedge
 {\cal D}\theta^{\mu}(-1)^{\epsilon_{\nu}}\,,
\end{eqnarray}
where
\beq
\label{g}
 g_{\mu\nu}=(-1)^{\epsilon_{\nu}\epsilon_{\mu}}g_{\nu\mu}\,,
 \qquad
 {\cal D}\theta^{\nu}=d\theta^{\nu}+
 \Gamma^{\nu}_{~i\lambda}(x)\,\theta^{\lambda}dx^i
 (-1)^{\epsilon_{\nu}+
 \epsilon_{\lambda}+\epsilon_i}\,,
 \qquad
 \epsilon(g_{\mu\nu})=\epsilon_{\mu}+\epsilon_{\nu}\,,
\eeq
while the curvature tensor is defined as follows
\beq
\label{R}
 R^{\nu}_{~\alpha ki}
 =-\Gamma^{\nu}_{~k\alpha ,i}(-1)^{\epsilon_k\epsilon_{\alpha}}+
 \Gamma^{\nu}_{~i\alpha ,k}(-1)^{\epsilon_i(\epsilon_k+\epsilon_{\alpha})}
 +\Gamma^{\nu}_{~k\beta}\Gamma^{\beta}_{~i\alpha}
 (-1)^{(\epsilon_k+\epsilon_i)\epsilon_{\alpha}}-
 \Gamma^{\nu}_{~i\beta}\Gamma^{\beta}_{~k\alpha}
 (-1)^{(\epsilon_i+\epsilon_k)\epsilon_{\alpha}+\epsilon_i\epsilon_k}\,.
\eeq
It is antisymmetric w.r.t. the last two indices:
 $R^{\nu}_{~\alpha i k}
 =-(-1)^{\epsilon_i\epsilon_k}R^{\nu}_{~\alpha k i}$.

The connection $\Gamma^\mu_{~k\nu}$ respects the metric
$g_{\mu\nu}$,
\begin{eqnarray}
\label{gc}
 g_{\mu\nu;k}=g_{\mu\nu,k}-g_{\mu\alpha}\Gamma^{\alpha}_{~k\nu}
 (-1)^{\epsilon_k\epsilon_{\nu}}-
 g_{\alpha\nu}\Gamma^{\alpha}_{~k\mu}
 (-1)^{\epsilon_{\nu}(\epsilon_{\alpha}+\epsilon_{\mu})+
 \epsilon_k\epsilon_{\mu}}=0\,,
\end{eqnarray}
and, under the change of coordinates,
\begin{eqnarray}
\label{LT}
 {\bar x}^i={\bar x}^i(x), \quad
 {\bar \theta}^{\nu}=\theta^{\mu}A_{\mu}^{\;\;\nu}(x),\quad
 \epsilon(A_{\mu}^{\;\;\nu})=\epsilon_{\mu}+\epsilon_{\nu}\,,
\end{eqnarray}
it transforms as follows:
\begin{eqnarray}
\label{Gtrn}
 {\bar \Gamma}^{\mu}_{~i\nu}
 =A^{\mu}_{\;\;\lambda}\Gamma^{\lambda}_{~k\alpha}
 \frac{\partial_r x^k}{\partial {\bar x}^i}B^{\alpha}_{\;\;\nu}
 (-1)^{\epsilon_{\alpha}(\epsilon_i+\epsilon_k)}-
 A^{\mu}_{\;\;\alpha ,k}B^{\alpha}_{\;\;\nu}
 \frac{\partial_r x^k}{\partial {\bar x}^i}
 (-1)^{\epsilon_{\alpha}\epsilon_k
 +\epsilon_{\nu}(\epsilon_i+\epsilon_k)}\,,
\end{eqnarray}
where
\begin{eqnarray}
\label{AB}
 A_{\mu}^{\;\;\nu}B_{\nu}^{\;\;\lambda}=
 B_{\mu}^{\;\;\nu}A_{\nu}^{\;\;\lambda}=\delta_{\mu}^{\lambda}\,,
 \qquad
 A_{\mu}^{\;\;\nu}=A^{\nu}_{\;\;\mu}
 (-1)^{\epsilon_{\mu}(\epsilon_{\nu}+1)}\,,
 \qquad
 B^{\mu}_{\;\;\nu}=B_{\nu}^{\;\;\mu}
 (-1)^{\epsilon_{\nu}(\epsilon_{\mu}+1)}\,.
\end{eqnarray}
Hence,  ${\cal D}\theta^{\nu}$ transforms homogeneous,
${\cal D}{\bar \theta}^{\nu}={\cal D}\theta^{\mu}A_{\mu}^{\;\;\nu}(x)$,
under the above change of coordinates (\ref{LT}).

Now, let us choose
\begin{eqnarray}
\label{gsp}
 g_{\mu\nu} = \omega_{ij}\,\epsilon_{ab}\,,
 \qquad
 \Gamma^{\nu}_{~k\mu}=  \Gamma^j_{~ ki}\,\delta^b_a\,.
\end{eqnarray}
Then the condition (\ref{gc}) takes the form
\begin{eqnarray}
\label{m}
 g_{\mu\nu;k}=0
 \quad \rightarrow \quad
 \epsilon_{ab}\,[\omega_{ij,k}-
 \omega_{il}\Gamma^l_{~kj}(-1)^{\epsilon_k\epsilon_j}
 +\omega_{jl}\Gamma^l_{~ki}
 (-1)^{\epsilon_i\epsilon_j+\epsilon_i\epsilon_k}]=0\,,
\end{eqnarray}
i.e.,~$\Gamma^j_{ki}$ in (\ref{gsp}) defines a symplectic
connection on the supermanifold ${\cal M}$ .
In this case we have the following representation for curvature
 tensor (\ref{R}) \cite{gl}:
\begin{eqnarray}
\label{Rsp0}
 R^{\alpha}_{~\nu kl}&=&\delta^a_b R^m_{~jkl}\,,
 \nonumber\\
 R^i_{\,jkl}&=&
 -\Gamma^i_{\,kj,l}
 + \Gamma^i_{\,lj,k}(-1)^{\epsilon_k\epsilon_l}
 +\Gamma^i_{\,km}\Gamma^m_{~lj}
  (-1)^{\epsilon_i+\epsilon_k(\epsilon_j+\epsilon_m)}
 -\Gamma^i_{\,lm}\Gamma^m_{~kj}
  (-1)^{\epsilon_i+\epsilon_l(\epsilon_j+\epsilon_m)+\epsilon_l\epsilon_k}\,,
\end{eqnarray}
with $R^i_{\,jkl}$ being the curvature of the supersymplectic
connection.

In  case when the connection is flat one is able to find a realization of the triplectic algebra on
${\cal M}$ \cite{gl}; for the general case this remains an open question.

\section{Superfield approach}

It seems to be advantageous to attack that problem in the more
general  superfield approach to $Sp(2)$ symmetric quantization \cite{l}.
In that approach we deal with the superfield $\phi^i$,
\beq
 \phi^i=x^i+\eta^a\theta^i_a+\eta^1\eta^2\; y^i\,,
 \quad{\rm where }\quad
 \theta^i_a =(-1)^{\epsilon_i}\omega^{ij}\theta_{ja}\,,
 \quad
 y^i=(-1)^{\epsilon_j}\omega^{ij}y_j\,,
\eeq
and  $\epsilon(y^i)=\epsilon_i$,
$\epsilon(\theta^i_a)=\epsilon_i+1$,
$\epsilon(\eta^a)=1$.

Let us  naively define, on the superfield space, the symplectic structure
$\omega(\phi)=\omega_{ij}(\phi)\,d\phi^j\wedge d\phi^i$.
Expanding this form  on Grassmann parameters  $\eta^a$, we obtain
\begin{eqnarray}
\label{wphi}
 \omega(\phi)=
 \omega_{ij}(x)\,dx^j\wedge dx^i - 2\eta^a\; dx^i\wedge d\theta_{ia}
 -\eta^1\eta^2
 \left[2dy_i\wedge dx^i + {\widetilde\Omega}(x,\theta)\right]\,,
\end{eqnarray}
where
\begin{eqnarray}
\label{W2}
 {\widetilde\Omega}=-\frac{1}{2}{\widetilde R}_{klij}(-1)^{\epsilon_l}
 \epsilon_{ba}\theta^{lb}\theta^{ka}
 dx^j\wedge dx^i +\omega_{ij}\epsilon_{ab}
 (-1)^{\epsilon_j}{\widetilde{\cal D}}\theta^{jb}\wedge
 {\widetilde{\cal D}}\theta^{ia}\,.
\end{eqnarray}
In (\ref{W2}) we used the notation:
\begin{eqnarray}
\label{Rw}
 {\widetilde R}_{klij}&=&\omega_{ij,kl}-\omega_{im,k}\omega^{mn}\omega_{nj,l}
 (-1)^{\epsilon_i+\epsilon_j+\epsilon_k+\epsilon_l+\epsilon_m+
 \epsilon_k(\epsilon_m+\epsilon_j)}+\\
 \nonumber
 &&\omega_{jm,k}\omega^{mn}\omega_{in,l}
 (-1)^{\epsilon_i+\epsilon_j+\epsilon_k+\epsilon_l+\epsilon_m+
 \epsilon_i\epsilon_j+\epsilon_k(\epsilon_m+\epsilon_i)}\,,\\
\label{DT}
 \widetilde{\cal D}\theta^{ia}&=&d\theta^{ia}
 +{\widetilde\Gamma}^i_{jk}\theta^{ka}dx^j
 (-1)^{(\epsilon_i+\epsilon_j+  \epsilon_k)}\,,
 \quad{\rm where}\quad
 {\widetilde\Gamma}^i_{jk}=\omega^{ip}\omega_{pj,k}
 (-1)^{\epsilon_j+\epsilon_k}\,.
\end{eqnarray}
Seemingly, the formal structure of ${\widetilde\Omega}$ is the same as for the supersymplectic structure (\ref{A}) in Section 2.
The quantity ${\widetilde R}_{ijkl}$ obeys the following symmetry properties,
\begin{eqnarray}
\label{Rws}
 {\widetilde R}_{klij}=-(-1)^{\epsilon_i\epsilon_j}{\widetilde R}_{klji}\,,
 \quad
 {\widetilde R}_{klij}=(-1)^{\epsilon_k\epsilon_l}{\widetilde R}_{lkij}\,,
\end{eqnarray}
as it should be for a supersymplectic curvature tensor. Of course, we also have to prove that ${\widetilde\Gamma}$ is a symplectic connection, i.e. respecting $\omega$, and that ${\widetilde R}$ may be rewritten in the form (\ref{Rsp0}). However, one can check that ${\widetilde\Gamma}^i_{jk}$
in (\ref{DT}) can not be considered as a symplectic connection
satisfying (\ref{m}).

The superfield symplectic structure $\omega(\phi)$
transforms covariant under a change of the superfield coordinates
${\bar \phi}^i={\bar\phi}^i(\phi)$:
\begin{eqnarray}
\label{Wt}
 {\bar\omega}({\bar\phi})=\omega(\phi),\quad
 d{\bar\phi}^i=d\phi^j\frac{\partial{\bar\phi}^i}{\partial\phi^j}\,,
 \quad
 {\bar\omega}_{ij}({\bar\phi})=\omega_{mn}(\phi)
 \frac{\partial_r\phi^n}{\partial{\bar\phi}^j}
 \frac{\partial_r\phi^m}{\partial{\bar\phi}^i}
 (-1)^{\epsilon_j(\epsilon_i+\epsilon_m)}\,.
\end{eqnarray}
In component form the coordinates transform according to
\begin{eqnarray}
\label{t}
 {\bar x}^i={\bar x}^i(x)\,,
 \quad
 {\bar\theta}^i_a=\theta^j_a\frac{\partial{\bar x}^i}{\partial x^j},
 \quad
 {\bar y}^i=y^j\frac{\partial{\bar x}^i}{\partial x^j}+
 \frac{1}{2}\epsilon_{ab}\theta^{ja}\theta^{kb}
 \frac{\partial^2{\bar x}^i}{\partial x^k\partial x^j}
 (-1)^{\epsilon_j}\,.
\end{eqnarray}
Hence, the even two-form $\omega(x)$, the pair of odd two-forms
$dx^i\wedge d\theta_{ia}$ as well as the two-form $2dy_i\wedge dx^i+{\widetilde\Omega}$
are covariant w.r.t. these transformations. Obviously, $\Omega$ itself
is not covariant under the given transformation, since
${\widetilde\Gamma}^i_{jk}$  is not a connection  
(and, therefore, ${\widetilde R}_{klij}$ (\ref{Rw}) is not a tensor) 
on ${\cal M}_0$.

Indeed, let us introduce, as in the previous section, the metric and the connection with components
\begin{eqnarray}
\label{g1}
 g_{\mu\nu}=\omega_{ij}\epsilon_{ab}\,,
 \qquad
 {\widetilde\Gamma}^{\mu}_{i\nu}
 =\widetilde\Gamma^{(j,a)}_{i(k,b)}
 =\widetilde\Gamma^j_{ik}\delta^a_b\,,
 \qquad \mu=(i,a)\,, \;\;\nu=(j,b)\,.
\end{eqnarray}
Notice, that ${\widetilde \Gamma}^i_{jk}$ does not respects the metric $g_{\mu\nu}$:
\begin{eqnarray}
\label{g2}
\nonumber
 g_{\mu\nu;k}\neq 0.
\end{eqnarray}
In principle, it is possible to achive respecting the metric if we change the definition of ${\widetilde \Gamma}^i_{jk}$ in (\ref{W2}) by omitting the factor $(-1)^{(\epsilon_i+\epsilon_j+  \epsilon_k)}$ in $\widetilde{\cal D}\theta^{ia}$.
But in any case ${\widetilde \Gamma}^i_{jk}$ does not transform,
under a change of local coordinates of the base supermanifold ${\cal M}_0$, $(x)\rightarrow ({\bar x})$, as a connection
\begin{eqnarray}
\label{Gn}
 {\bar \Gamma}^i_{jk} \neq
 \frac{\partial_r {\bar x}^i}{\partial x^p}
 \Gamma^p_{mn}\frac{\partial_r x^n}{\partial {\bar x}^k}
 \frac{\partial_r x^m}{\partial {\bar x}^j}
 (-1)^{\epsilon_k(\epsilon_j+\epsilon_m)}+
 \frac{\partial_r {\bar x}^i}{\partial x^p}
 \frac{\partial^2_r x^p}{\partial {\bar x}^j \partial {\bar x^k}}.
\end{eqnarray}
Hence, {\it $\widetilde \Gamma^i_{jk}$ could not be interpreted as a 
connection on ${\cal M}_0$}!
Similarly, ${\widetilde R}_{ijkl}$ could not be considered as a curvature
of the connection on ${\cal M}_0$
This explains, why  the two-form $\Omega$ is not covariant under the transformation ${\bar x}^i={\bar x}^i(x)$,
${\bar\theta}^i_a=\theta^j_a\frac{\partial{\bar x}^i}{\partial x^j}$.

\medskip

{\sc Acknowledgements:}~
A.N. would like to thank the Institute of Theoretical Physics of the Leipzig University for kind hospitality and the Graduate College ``Quantum Field Theory'' for financial support, which made possible his participation in
presented  research. The work of B.G. and P.L was supported by
Deutsche Forschungsgemeinschaft, grant DFG 436 RUS 17/15/04.
We also acknowledge partial support by RFBR grant
03-02-16193 and the President grant 1252.2003.2 (P.L.) as well as
INTAS grants 03-51-6346 (P.L) and 00-00262 (A.N).

\end{document}